\documentclass[12pt,draftcls,onecolumn]{IEEEtran}
\usepackage{cite}
\usepackage{amsfonts,amssymb}
\usepackage{amsbsy}
\usepackage{graphicx}
\usepackage{latexsym}
\usepackage{amsmath,mathrsfs}
\usepackage{psfrag}
\usepackage{subfigure}
\newtheorem{Def}{Definition}
\newtheorem{Eg}{Example}
\newtheorem{Prop}{Proposition}
\newtheorem{Thm}{Theorem}
\newtheorem{Lem}{Lemma}
\newtheorem{Coro}{Corollary}
\newtheorem{Rem}{Remark}

\newcommand{\ld}{\lambda}

\begin{document}
\title{Similarity-Based Supervisory Control of Discrete Event Systems}
\author{Yongzhi Cao and Mingsheng Ying%
\thanks{This work was supported by the National Foundation of
Natural Sciences of China under Grants 60496321, 60321002, and
60273003 and by the Key Grant Project of Chinese Ministry of
Education.}%
\thanks{The authors are with the State Key Laboratory of Intelligent Technology and
Systems, Department of Computer Science and Technology, Tsinghua
University, Beijing 100084, China (e-mail:
caoyz@mail.tsinghua.edu.cn, yingmsh@mail.tsinghua.edu.cn).}}

\maketitle
\begin{abstract}Due to the appearance of uncontrollable events in discrete event systems, one may wish to
replace the behavior leading to the uncontrollability of
pre-specified language by some quite similar one. To capture this
similarity, we introduce metric to traditional supervisory control
theory and generalize the concept of original controllability to
$\ld$-controllability, where $\ld$ indicates the similarity degree
of two languages. A necessary and sufficient condition for a
language to be $\ld$-controllable is provided. We then examine
some properties of $\ld$-controllable languages and present an
approach to optimizing a realization.
\end{abstract}

\section{Introduction}
\PARstart{S}{upervisory} control theory (SCT) initiated by Ramadge
and Wonham \cite{RW87} and subsequently extended by other
researchers (see, for example, \cite{RW89,CL99} and the
bibliographies therein) provides a systematic approach to
controlling discrete event systems (DES). The behavior of a DES is
represented by a language over the set of events, and in the
paradigm of standard SCT, Ramadge and Wonham \cite{RW87} have
formulated supervisory control problem by two languages that
correspond to minimal acceptable behavior and legal behavior,
respectively. In this formulation, both general and nonblocking
solutions are well discussed.

Because of some practical requirements of control engineering, the
standard SCT has been extended from the aspect of control
objective. It has been observed by Lafortune and Chen \cite{LC90}
that the control objective of requiring nonblocking solutions is
too conservative in some cases, and thus they have developed the
supervisory control problem with blocking in \cite{LC90,CL91}.
Subsequently, Lafortune and Lin \cite{LL90,LL91} formulated and
solved a more general supervisory control problem whose control
objective is given by ``desired" behavior and ``tolerated"
behavior. The motivation behind this is that one can achieve more
desired behavior by tolerating some behavior that will exceed the
ideal desired one. Using probability to precisely specify what is
tolerable was first presented by Lin in \cite{L90}. This work was
further developed in probabilistic DES \cite{LLL99}. The research
mentioned above shows that to achieve more desired behavior,
sometimes it is worth tolerating undesirable behavior especially
in some systems whose constraints are not rigid.

\markboth{CAO AND YING: Similarity-Based Supervisory Control of
Discrete Event Systems}{CAO AND YING: Similarity-Based Supervisory
Control of Discrete Event Systems}

Tolerable behavior, which depends on different practical systems,
gives rise to different supervisory control problems. In this
paper, we are interested in the supervisory control problem in
which one can accept some behavior quite similar to desired one.
This is motivated by the fact that some similar behavior often
occurs in some DES and one may wish to tolerate similar behavior
when the ideal desired one is not feasible. For example, assume
that in a common computer system, the jobs completed by CPU
(central processing unit) will request access to peripheral
devices consisting of one printer and one disk. It seems
reasonable to expect that if the default device is busy or wrong,
the jobs will give access to the other device.

In order to capture the similarity of behavior, we first suppose
that the event set of a DES is equipped with a metric $d$. This
hypothesis is not too constrained since any nonempty set can be
endowed with at least the discrete metric. The metric $d$
indicates the similarity of events. Based upon this metric, we
then construct a distance function $\tilde{d}$ for all pairs of
event strings by using so-called Baire metric. Finally, the
Hausdorff metric $\tilde{d}_H$ induced by $\tilde{d}$ can serve as
a similarity measure on the set of languages. The less the value
of $\tilde{d}_H$, the more similar the two languages. With this
similarity measure, we propose the concept of
$\ld$-controllability, where $\ld$ stands for similarity index.
More explicitly, we say that a language $K$ is $\ld$-{\it
controllable} if there exists a controllable language
$\widetilde{K}$ satisfying that
$\tilde{d}_H(\widetilde{K},K)\leq\ld$. Such a $\widetilde{K}$ is
called a {\it realization} of $K$. Clearly, each controllable
language in the sense of SCT is $\ld$-controllable, and moreover,
$0$-controllability coincides with original controllability.
Hence, the notion of $\ld$-controllability is a generalization of
the original controllability in SCT. In some applications, the
specifications offered by users may be relaxed. If a specification
is not controllable and some dissimilarities between events can be
tolerated, then we can turn our attention to finding a similar one
by using $\ld$-controllability, which increases the intelligence
of standard supervisory control.

In our setting, we still use the traditional supervisor to control
the system; the control objective which is different from the
aforementioned ones is, however, to find a realization of the
pre-specified desired language. In other words, the control
objective here is to achieve certain behavior similar to the
desired one. Taking similarity of elements into account and using
metric to describe the similarity are widely recognized in some
fields of Computer Science such as metric semantics, process
calculus, and pattern recognition (see, for example,
\cite{dBdV96}, \cite{vB98}, \cite{TK03}). In the earlier work
\cite{R89}, a distance function defined in \cite{E74} is also used
to characterize the infinite or sequential behavior of DES, and
moreover, a generalized notion of controllability for
$\omega$-languages is introduced. Such a notion essentially
depends on the prefix of $\omega$-language under consideration,
and thus it cannot serve our purpose of similarity-based
supervisory control. Recently, a signed real measure for
sublanguages of regular languages has been formulated and studied
in \cite{RP03,RSP03}. The measure which is different from our
similarity measure only serves as an evaluation of supervisors.
Perhaps there is a deep connection between them, and this is an
interesting problem for the future study. Related to the metric
for events, in Petri nets the synchronic distance between
transitions has been introduced by Petri \cite{P76} to describe
the degree of mutual dependence between events in a
condition/event system (see \cite{M89} and the bibliographies
therein for further information on synchronic distances).

The purpose of this paper is to introduce the idea of
similarity-based supervisory control, and we only concentrate on
some basic aspects of $\ld$-controllability. We first examine some
algebraic properties of $\ld$-controllable languages, and then
present a necessary and sufficient condition for a language to be
$\ld$-controllable. An algorithm for determining whether or not a
finite language is $\ld$-controllable is also provided. Further,
we show that the supremal $\ld$-controllable sublanguage of a
given language exists, and discuss some of its properties.
Finally, for a given $\ld$-controllable language $K$, we turn our
attention to finding a Pareto optimal realization $\widetilde{K}$
in the sense that it is impossible to enlarge the common behavior
$\widetilde{K}\cap K$ and simultaneously reduce the different
behavior $\widetilde{K}\backslash K$.

The rest of the paper is organized as follows. In Section II, we
review some basics of SCT and metric space. In Section III, we
introduce the concept of $\ld$-controllability, discuss some
properties of $\ld$-controllable languages, and present a
necessary and sufficient condition for a language to be
$\ld$-controllable. The supremal $\ld$-controllable sublanguage is
addressed in this section as well. Section IV is devoted to
deriving a Pareto optimal realization from an arbitrary
realization. We provide an illustrative example in Section V and
conclude the paper in Section VI.

\section{Preliminaries}
Let $E$ denote the finite set of events, and $E^*$ denote the set
of all finite sequences of events, or stings, in $E$, including
the empty string $\epsilon$. The length of a string $\omega$ is
denoted by $l(\omega)$, and  the prefix closure of a language $L$
is denoted by $\overline{L}$.

The DES to be controlled is modelled by a deterministic automaton:
$G=(Q,E,\delta,q_0)$, where $Q$ is a set of states with the
initial state $q_0$, $E$ is a set of events, and $\delta:Q\times
E\rightarrow Q$ is a (partial) transition function. The function
$\delta$ is extended to $\delta:Q\times E^*\rightarrow Q$ in the
obvious way. The behavior of a DES is modelled as a prefix closed
language $L(G)=\{s\in E^*:\delta(q_0,s)\mbox{ is defined}\}$.

The supervisory control theory partitions the event set into two
disjoint sets of controllable and uncontrollable events, $E_c$ and
$E_{uc}$, respectively. A supervisor is a map
$S:L(G)\rightarrow2^E$ such that $S(s)\supseteq E_{uc}$ for any
string $s\in L(G)$. The language generated by the controlled
system is denoted by $L(S/G)$. Following \cite{RW87}, a language
$K\subseteq L(G)$ is said to be {\it controllable} (with respect
to $L(G)$ and $E_{uc}$) if $\overline{K}E_{uc}\cap
L(G)\subseteq\overline{K}.$ It has been shown in \cite{RW87} that
a given nonempty language $K\subseteq L(G)$ is controllable if and
only if there exists a supervisor $S$ such that
$L(S/G)=\overline{K}$.

For any language $K$, there exist the supremal controllable
sublanguage \cite{RW87} and the infimal prefix closed and
controllable superlanguage \cite{LC90} of $K$, denoted by
$K^\uparrow$ and $K^\downarrow$, respectively. For more details
about the theory of DES, we refer the reader to, for example,
\cite{CL99}.

Let us collect some basic notions on metric space.

\begin{Def}
A {\it ($1$-bounded) metric space} is a pair $(X,d)$ consisting of
a nonempty set $X$ and a function $d:X\times X\longrightarrow
[0,1]$ which satisfies the following conditions:

(M1) $d(x,y)=0$ if and only if $x=y$,

(M2) $d(x,y)=d(y,x)$ for all $x,y\in X$, and

(M3) $d(x,z)\leq d(x,y)+d(y,z)$ for all $x,y,z\in X$.\label{Dd}
\end{Def}

The distance $d(x,y)$ measures the similarity between $x$ and $y$.
The less the distance, the more similar the two elements. To
simplify notation, sometimes we write $X$ instead of $(X,d)$.
Recall that if $(X,d)$ is a metric space and $M\subseteq X$, then
 $(M,d|_{M\times M})$ is also a metric
space, where $d|_{M\times M}$ is the restriction of $d$ to $M$.

Let $(X,d)$ be a metric space, $x_0\in X$, and $\ld\in[0,1]$. The
set $B(x_0,\ld)=\{x\in X: d(x_0,x)\leq\ld\}$ is called the {\it
$\ld$-ball about $x_0$}; for a subset $A$ of $X$, by the {\it
$\ld$-ball about $A$} we mean that the set $B(A,\ld)=\cup_{x\in
A}B(x,\ld)$. We extend $d$ to a pair $x,A$, where $x\in X$ and
$A\subseteq X$, by defining $d(x,A)=\inf\limits_{a\in A}d(x,a)$ if
$A\neq\emptyset$, and $d(x,A)=1$ otherwise. Further, we define
{\it Hausdorff metric} for a pair $A,B\subseteq X$ as follows:
\begin{displaymath}
d_H(A,B)=\left\{ \begin{array}{ll}
0, \hfill\textrm{ if}&\!\!\!A=B=\emptyset\\
\max\{\sup\limits_{a\in A}d(a,B),\sup\limits_{b\in B}d(b,A)\},
&\textrm{otherwise.}
\end{array} \right.
\end{displaymath}
The Hausdorff metric is one of the common ways of measuring
resemblance between two sets in a metric space; it satisfies the
conditions (M2) and (M3) in Definition \ref{Dd}, but it does not
satisfy the condition (M1) in general.

\section{Metric Controllability}
Let us begin with the (finite) event set $E$ of a DES $G$ and a
metric $d$ on $E$ which makes $E$ into a metric space. We now
endow $E^{*}$ with the {\it Baire metric} induced by $d$, which
measures the distance between strings and pays more attention to
the events occurring antecedently. Let $s=s_1s_2\cdots s_{l(s)}$
and $t=t_1t_2\cdots t_{l(t)}$ be two strings in $E^*$, and
$l(s,t)=\max\{l(s),l(t)\}$. If $l(s)\neq l(t)$, say $l(s)<l(t)$,
we take $s_i=\epsilon$ for each $i>l(s)$. We then define
$$\tilde{d}(s,t)=\sum_{i=1}^{l(s,t)}\frac{1}{2^i}d(s_i,t_i),$$ where we set
$d(\epsilon,\epsilon)=0,$ and $d(a,\epsilon)=d(\epsilon,a)=1$ for
any $a\in E$. It is easy to verify that $\tilde{d}$ does give rise
to a metric on $E^*$. For later need, we make a useful
observation.
\begin{Lem}Let $L\subseteq L(G)$ and $s\in L(G)$. Then $\inf\limits_{t\in L}\tilde{d}(s,t)=\min\limits_{t\in
L}\tilde{d}(s,t)$, namely, $\tilde{d}(s,L)=\min\limits_{t\in
L}\tilde{d}(s,t)$.\label{Lmin}
\end{Lem}
\begin{proof}Set $W=\{w\in \overline{L}:l(w)\leq l(s)\}$. It is a finite set since the event set
$E$ is finite. For each $w\in W$, we choose a string $w'$
satisfying the following:
\begin{itemize}
    \item $w'=wv'\in L$, where $v'\in E^*$; and
    \item if $wv''\in L$ for some $v''\in E^*$, then $l(v'')\geq l(v')$.
\end{itemize}
It follows from the definition of $W$ that such a $w'$ does exist,
but it may not be unique. It does not matter since we need only
one representative of them. Let $W'$ be the set consisting of all
such $w'$. Then the cardinality of $W'$ is less than or equal to
that of $W$. Further, set $L'=\{w\in L:l(w)<l(s)\}\cup W'$.
Clearly, $L'$ is a finite set, so $\min\limits_{w\in
L'}\tilde{d}(s,w)$ exists.

For any $t\in L$, we claim that
$\tilde{d}(s,t)\geq\min\limits_{w\in L'}\tilde{d}(s,w)$. In fact,
for the case that $l(t)\leq l(s)$, we have that $t\in L'$. Hence
$\tilde{d}(s,t)\geq\min\limits_{w\in L'}\tilde{d}(s,w)$. In the
other case that $l(t)>l(s)$, we can write $t$ as $w_tv_t$
satisfying that $l(w_t)=l(s)$. If $w_tv_t\in W'$, then it is clear
that $\tilde{d}(s,t)\geq\min\limits_{w\in L'}\tilde{d}(s,w)$;
otherwise, by the definition of $W'$ there exists $v'_t\in E^*$
with $l(v'_t)\leq l(v_t)$ such that $w_tv'_t\in W'$. We thus get
by the definition of Baire metric that
$$\tilde{d}(s,t)=\tilde{d}(s,w_tv_t)\geq\tilde{d}(s,w_tv'_t)\geq\min\limits_{w\in L'}\tilde{d}(s,w).$$
Therefore the claim holds. Note that $L'\subseteq L$, hence
$\min\limits_{t\in L}\tilde{d}(s,t)=\min\limits_{w\in
L'}\tilde{d}(s,w)$, and thus $\inf\limits_{t\in
L}\tilde{d}(s,t)=\min\limits_{t\in L}\tilde{d}(s,t)$, as desired.
\end{proof}

\vspace{0.3cm}As mentioned earlier, Hausdorff metric does not give
rise to a metric space in general. However, if we consider the
powerset $\mathcal{P}(E^*)$ of $E^*$ with the Hausdorff metric
induced by $\tilde{d}$, then we can get a metric space.

\begin{Prop}Let $\tilde{d}_H$ be the Hausdorff metric induced by
the metric $\tilde{d}$ introduced above. Then
$(\mathcal{P}(E^*),\tilde{d}_H)$ is a metric space.
\end{Prop}
\begin{proof}As mentioned earlier, any Hausdorff metric satisfies the conditions (M2) and (M3) in Definition \ref{Dd},
so we only need to check the condition (M1). Suppose that
$\tilde{d}_H(A,B)=0$, where $A,B\subseteq E^*$. Seeking a
contradiction, assume that $A\neq B$; without loss of generality,
we may assume that there exists $s\in A\backslash B$. By the
definition of Hausdorff metric, we know from $\tilde{d}_H(A,B)=0$
that $\tilde{d}(s,B)=0$. This means that $B\neq\emptyset$, and
moreover, $\min\limits_{t\in B}\tilde{d}(s,t)=0$ by Lemma
\ref{Lmin}. Since $\tilde{d}$ is a metric on $E^*$, the latter
forces that $s\in B$, a contradiction. We thus get that $A=B$.
Conversely, if $A=B$, then it is obvious that
$\tilde{d}_H(A,B)=0$. So $\tilde{d}_H$ is a metric on
$\mathcal{P}(E^*)$, thus finishing the proof.
\end{proof}

\vspace{0.3cm}The Hausdorff metric defined above measures the
similarity of two languages. For convenience of notation, we will
write $d$ for the metrics $\tilde{d}$ and $\tilde{d}_H$ induced by
$d$ in what follows; it will be always clear from the context
which metric is being considered. As a subset of $E^*$, $L(G)$ is
a metric space with restricted metric. From now on, we will work
in $L(G)$ instead of $E^*$, unless otherwise specified. We can now
introduce the key notion.

\begin{Def}Given $\ld\in[0,1]$, a language $K\subseteq L(G)$ is said
to be {\it $\ld$-controllable} (with respect to $L(G)$ and
$E_{uc}$) if there exists a language $\widetilde{K}\subseteq L(G)$
satisfying the following conditions:

1) \ $d(\widetilde{K},K)\leq\ld$;

2) \ $\widetilde{K}$ is controllable with respect to $L(G)$ and
$E_{uc}$. \\ If such a $\widetilde{K}$ exists, we call it a {\it
realization} of $K$.
\end{Def}

Intuitively, a language $K$ is $\ld$-controllable if there is a
controllable language that is similar to $K$. Observe that each
controllable language is $\ld$-controllable. The following
example, however, shows that the converse is not true in general.
\begin{Eg}Let $L(G)=\{\epsilon,a,ab,ag,af,abc,age\}$, $E_{uc}=\{c,g,f\}$, and $K=\{\epsilon,a,ab,af\}$. It is
easy to see that $K$ is not controllable. Let us now define a
metric $d$ on $E$ as follows:
\begin{displaymath}
d(x,y)=\left\{ \begin{array}{ll} 0, &\textrm{if $x=y$}\\
0.01, & \textrm{if $(x,y)=(b,g)$ or $(g,b)$}\\
1, & \textrm{otherwise.}
\end{array} \right.
\end{displaymath}Based on this metric, we can obtain the induced metrics on $L(G)$ and $\mathcal{P}(L(G))$,
respectively. For example, $d(ab,ag)=0.0025$ and
$d(K,\{\epsilon,a,ag,af\})=0.0025$. Observe that
$\{\epsilon,a,ag,af\}$ is controllable and it can serve as a
realization of $K$ whenever $\ld\geq0.0025$. Therefore, according
to our definition, $K$ is $\ld$-controllable for any
$\ld\geq0.0025$.\label{Eclo}
\end{Eg}

Let us give some remarks on the concept of $\ld$-controllability.
\begin{Rem}\

1) \ A language $K$ is $0$-controllable if and only if $K$ is
controllable. Note also that one can endow any event set $E$ with
discrete metric and educe further Hausdorff
 metric on $\mathcal{P}(E^*)$. Thus in view of this, the concept of $\ld$-controllability is also a generalization of the
 ordinary controllability in the framework of Ramadge-Wonham.

2) \ If $K$ is $\ld_1$-controllable and $\ld_1\leq\ld_2$, then $K$
is also $\ld_2$-controllable. In particular, each controllable
language is $\ld$-controllable, for any $\ld\in[0,1]$.

3)\ If $K$ is $\ld$-controllable, then so is $\overline{K}$. But
the converse does not hold in general.\label{Pcol}
\end{Rem}
{\it Proof of 3):} Let $\widetilde{K}$ be a realization of $K$. We
want to show that $\overline{\widetilde{K}}$ is a realization of
$\overline{K}$. Since $\overline{\widetilde{K}}$ is controllable
by definition, it suffices to verify that
$d(\overline{\widetilde{K}},\overline{K})\leq\ld$, namely,
$\sup\limits_{s\in\overline{K}}d(s,\overline{\widetilde{K}})\leq\ld$
and
$\sup\limits_{s\in\overline{\widetilde{K}}}d(s,\overline{K})\leq\ld$.
By definition and Lemma \ref{Lmin}, the former is equivalent to
$\min\limits_{t\in \overline{\widetilde{K}}}d(s,t)\leq\ld$ for any
$s\in\overline{K}$, while the latter is equivalent to
 $\min\limits_{t\in \overline{K}}d(s,t)\leq\ld$ for any
$s\in\overline{\widetilde{K}}$.  We only prove the former; the
latter can be proved in a similar way. Let $s$ be an arbitrary
string in $\overline{K}$. Then there exists $s'\in E^*$ satisfying
that $ss'\in K$. As $\widetilde{K}$ is a realization of $K$, we
have that $\sup\limits_{w\in K}d(w,\widetilde{K})\leq\ld$, that
is, $\min\limits_{v\in \widetilde{K}}d(w,v)\leq\ld$ for any $w\in
K$. In particular, setting $w=ss'$, we have at least one
$v\in\widetilde{K}$ such that $d(ss',v)\leq\ld$. If $l(v)\geq
l(s)$, then take $t$ to be the prefix of $v$ with length $l(s)$;
otherwise, take $t=v$. Clearly, such a selection of $t$ satisfies
that $t\in\overline{\widetilde{K}}$ and yields that $d(s,t)\leq
d(ss',v)\leq\ld$ by the definition of Baire metric. Therefore,
$\min\limits_{t\in \overline{\widetilde{K}}}d(s,t)\leq\ld$. As
$s\in\overline{K}$ was arbitrary, this completes the proof of the
first part.

For the second part, consider the following example: Let
$L(G)=\{\epsilon,a,c,ab\}$, $E_{uc}=\{c\}$, $K=\{ab\}$, and $d$ be
a metric defined on $E=\{a,b,c\}$ as follows:
\begin{displaymath}
d(x,y)=\left\{ \begin{array}{ll} 0, &\textrm{if $x=y$}\\
0.02, & \textrm{if $(x,y)=(a,c)$ or $(c,a)$}\\
1, & \textrm{otherwise.}
\end{array} \right.
\end{displaymath}
Set $\ld=0.01$. Then $\overline{K}=\{\epsilon,a,ab\}$ is
$0.01$-controllable since $L(G)$ can serve as its realization.
Nevertheless, there is no $K'\subseteq L(G)$ satisfying that
${K'}$ is both controllable and $d(K',K)\leq0.01$. So $K$ is not
$0.01$-controllable.\hfill$\blacksquare$

\vspace{0.3cm}The following are some useful properties of
$\ld$-controllable languages.
\begin{Prop}\

1) \ If $K_1$ and $K_2$ are $\ld$-controllable, then so is
$K_1\cup K_2$.

2) \ If $K_1$ and $K_2$ are $\ld$-controllable, then $K_1\cap K_2$
need not be $\ld$-controllable.

3) \ If $\widetilde{K}_1$ and $\widetilde{K}_2$ are two
realizations of $K$, then so is
$\widetilde{K}_1\cup\widetilde{K}_2$. But
$\widetilde{K}_1\cap\widetilde{K}_2$ is not necessarily a
realization of $K$.\label{PContr}
\end{Prop}
\begin{proof}\

1) Suppose that $\widetilde{K}_i$, $i=1,2$, is a realization of
$K_i$. It is easy to verify that
$\widetilde{K}_1\cup\widetilde{K}_2$ is a realization of $K_1\cup
K_2$.

2) Consider the following counter example: Keep $L(G), E_{uc}$,
and $d$ in Example \ref{Eclo}. Let $K_1=\{\epsilon,a,ab,af\}$,
$K_2=\{\epsilon,a,ag,af\}$, and $\ld=0.0025$. Then $K_2$ is
controllable, and moreover, $d(K_1,K_2)=0.0025$. Hence both $K_1$
and $K_2$ are $0.0025$-controllable. But, by a simple computation,
one can find that $K_1\cap K_2=\{\epsilon,a,af\}$ is not
$0.0025$-controllable.

3) The first part follows immediately from the definition of
$\ld$-controllability. For the second part, one can easily give a
counter example. In fact, there is one at the end of Section V,
where the intersection of two Pareto optimal realizations of $K$
fails to be a realization.
\end{proof}

\vspace{0.3cm}
The following theorem presents a necessary and sufficient
condition for a language $K\subseteq L(G)$ to be
$\ld$-controllable via its $\ld$-ball in $L(G)$.
\begin{Thm}A language $K\subseteq L(G)$ is $\ld$-controllable if and only if
$d(B(K,\ld)^{\uparrow},K)\leq\ld$.\label{TContr}
\end{Thm}
\begin{proof}
The sufficiency follows immediately from the definition, so we
only need to prove the necessity. Suppose that $K$ is
$\ld$-controllable. By definition, there exists a controllable
language $\widetilde{K}\subseteq L(G)$ with
$d(\widetilde{K},K)\leq\ld$. It follows that $d(x,K)\leq\ld$ for
any $x\in\widetilde{K}$, that is, $\min\limits_{y\in
K}d(x,y)\leq\ld$ by Lemma \ref{Lmin}. Therefore there exists
$y_x\in K$ such that $d(x,y_x)\leq\ld$, and thus we get that $x\in
B(y_x,\ld)\subseteq B(K,\ld)$ for any $x\in\widetilde{K}$. It
means that $\widetilde{K}\subseteq B(K,\ld)$, and furthermore,
$\widetilde{K}\subseteq B(K,\ld)^{\uparrow}$, which implies that
$d(s,B(K,\ld)^{\uparrow})\leq\ld$ for any $s\in K$. Consequently,
$\sup\limits_{s\in K}d(s,B(K,\ld)^{\uparrow})\leq\ld$. On the
other hand, since $B(K,\ld)^{\uparrow}\subseteq B(K,\ld)$, we have
that $d(t,K)\leq\ld$ for any $t\in B(K,\ld)^{\uparrow}$, which
yields that $\sup\limits_{t\in B(K,\ld)^{\uparrow}}d(t,K)\leq\ld$.
Hence, we obtain that $d(B(K,\ld)^{\uparrow},K)\leq\ld$. The proof
is completed.
\end{proof}

We would like to develop an algorithm for determining whether a
finite language is $\ld$-controllable. For this purpose, we need
an algorithm for computing $\ld$-ball about a string.

Let $G=(Q,E,\delta,q_0)$ be a deterministic automaton and
$s=s_1s_2\cdots s_n$ be a fixed string in $L(G)$. Let $d$ be the
metric on $L(G)$ defined as before.
 For each $q\in Q$, define $E(q)=\{e\in
E:\delta(q,e)\mbox{ is defined}\}$. Recall that $B(s,\ld)=\{r\in
L(G):d(r,s)\leq\ld\}$.

{\tt {\it Algorithm for $B(s,\ld)$:}

\hspace{0.3cm} $B(s,\ld)\leftarrow \emptyset$;

\hspace{0.3cm} $r\leftarrow \epsilon$;

\hspace{0.3cm} $i\leftarrow 1$;

\hspace{0.3cm} $n\leftarrow l(s)$;

\hspace{0.3cm} $F(\ld,r,i)$;

end Algorithm for $B(s,\ld)$.

Here the procedure $F(\ld,r,i)$ is defined recursively as follows:

{\it Procedure $F(\ld,r,i)$:}

\hspace{0.3cm}if $i=n+1$ then

\hspace{0.6cm}if $\ld\geq \frac{1}{2^n}$ then

\hspace{0.9cm} $B(s,\ld)\leftarrow B(s,\ld)\cup \{rr'\in
L(G):r'\in E^*$\};

\hspace{0.6cm}else

\hspace{0.9cm} $k\leftarrow \lfloor-\log_2(1-2^n\ld)\rfloor$;

\hspace{0.9cm} $B(s,\ld)\leftarrow B(s,\ld)\cup \{rr'\in L(G):
l(r')\leq k\}$;

\hspace{0.6cm}end if

\hspace{0.6cm}return;

\hspace{0.3cm}end if

\hspace{0.3cm}if $\ld\geq \frac{1}{2^i}+\cdots+\frac{1}{2^n}$ then

\hspace{0.6cm}$B(s,\ld)\leftarrow B(s,\ld)\cup \{r\}$;

\hspace{0.3cm}end if

\hspace{0.3cm}for each $e\in E(\delta(q_0,r))$

\hspace{0.6cm}if $\ld\geq\frac{d(s_i,e)}{2^i}$ then

\hspace{0.9cm}$r'\leftarrow re$;

\hspace{0.9cm}$\ld'\leftarrow\ld-\frac{d(s_i,e)}{2^i}$;

\hspace{0.9cm}$i'\leftarrow i+1$;

\hspace{0.9cm}$F(\ld',r',i')$;

\hspace{0.6cm}end if

\hspace{0.3cm}end for

end Procedure $F(\ld,r,i)$.}\hfill$\blacksquare$

The correctness of the above algorithm follows directly from the
definition of Baire metric. The worst-case complexity of
calculating $B(s,\ld)$ is $O(|E|^{l(s)})$.

Based on the above algorithm and Theorem \ref{TContr}, we are now
ready to provide  an algorithm for determining whether a finite
language is $\ld$-controllable. Notice that
$B(K,\ld)^{\uparrow}\subseteq B(K,\ld)$, so by definition we have
that the condition $d(B(K,\ld)^{\uparrow},K)\leq\ld$ in Theorem
\ref{TContr} holds if and only if $\sup\limits_{s\in
K}d(s,B(K,\ld)^{\uparrow})\leq\ld$. By Lemma \ref{Lmin}, the
latter is equivalent to that $\min\limits_{b\in
B(K,\ld)^{\uparrow}}d(s,b)\leq\ld$ for any $s\in K$. Further, this
is equivalent to that $B(s,\ld)\cap
B(K,\ld)^{\uparrow}\neq\emptyset$ for any $s\in K$. Note that
$B(K,\ld)=\cup_{s\in K}B(s,\ld)$ and one can compute
$B(K,\ld)^{\uparrow}$ by using standard algorithm for the
operation ``$\uparrow$" developed in
\cite{WR87},\cite{BGK90,LC90}. Summarily, we have the following
result.

\begin{Thm}To decide whether or not a finite language $K$ is $\ld$-controllable,
we can follow the steps below:
\begin{enumerate}
    \item For all $s\in K$, compute $B(s,\ld)$ by using {\it Algorithm for
    $B(s,\ld)$}.
    \item Compute $B(K,\ld)^{\uparrow}$ by using standard
algorithm for the operation ``$\uparrow$".
    \item Decide whether or not each $s\in K$ satisfies that $B(s,\ld)\cap
    B(K,\ld)^{\uparrow}\neq\emptyset$.
\end{enumerate}
If there exists $s\in K$ such that $B(s,\ld)\cap
B(K,\ld)^{\uparrow}=\emptyset$, then $K$ is not
$\ld$-controllable; otherwise, $K$ is $\ld$-controllable.
\end{Thm}

From 3) of Proposition \ref{PContr}, we see that the union of two
realizations of a $\ld$-controllable language $K$ is still a
realization. This can be easily generalized to infinite unions and
gives rise to the supremal realization of $K$. The next
observation shows us the relationship between the supremal
realization of $K$ and the $\ld$-ball about $K$.

\begin{Prop}
Let $K$ be a $\ld$-controllable language and $\widetilde{K}_i,i\in
I$, be all realizations of $K$. Then $\cup_{i\in
I}\widetilde{K}_i=B(K,\ld)^{\uparrow}$.\label{PRea}
\end{Prop}
\begin{proof}We know by Theorem \ref{TContr} that $B(K,\ld)^{\uparrow}$ is a realization of $K$, so
$B(K,\ld)^{\uparrow}\subseteq\cup_{i\in I}\widetilde{K}_i$.
Conversely, since $d(\widetilde{K}_i,K)\leq\ld$ for each $i\in I$,
it follows that $d(x,K)\leq\ld$ for any $x\in\widetilde{K}_i$,
that is, $\min\limits_{y\in K}d(x,y)\leq\ld$ by Lemma \ref{Lmin}.
Therefore there exists $y_x\in K$ such that $d(x,y_x)\leq\ld$, and
thus we have that $x\in B(y_x,\ld)\subseteq B(K,\ld)$ for any
$x\in\widetilde{K}_i$. It means that $\widetilde{K}_i\subseteq
B(K,\ld)$, and we get that $\cup_{i\in I}\widetilde{K}_i\subseteq
B(K,\ld)$. Note that $\cup_{i\in I}\widetilde{K}_i$ is
controllable, hence we have that $\cup_{i\in
I}\widetilde{K}_i\subseteq B(K,\ld)^{\uparrow}$, finishing the
proof of the proposition.
\end{proof}

\vspace{0.3cm}We end this section with a discussion on supremal
$\ld$-controllable sublanguage. To this end, let us define the
class of $\ld$-controllable sublanguages of $K$ as follows:
$$\mathscr{C}_\ld(K)=\{M\subseteq K:M\mbox{ is $\ld$-controllable}\}.$$
Observe that $\emptyset\in\mathscr{C}_\ld(K)$, so the class is not
empty. Define $K^\Uparrow=\cup_{M\in\mathscr{C}_\ld(K)}M$. Note
that 1) of Proposition \ref{PContr} can be easily generalized to
infinite unions, hence $K^\Uparrow$ gives rise to the largest
$\ld$-controllable sublanguage of $K$, where ``largest" is in
terms of set inclusion. We call $K^\Uparrow$ the {\it supremal
$\ld$-controllable sublanguage} of $K$ and refer to ``$\Uparrow$"
as the operation of obtaining the supremal $\ld$-controllable
sublanguage. Clearly, $K^\uparrow\subseteq K^\Uparrow$. If $K$ is
$\ld$-controllable, then $K^\Uparrow=K$. In the ``worst" case,
$K^\Uparrow=\emptyset$.

Several useful properties of the operation are presented in the
following proposition.

\begin{Prop}\

1) \ If $K$ is prefix closed, then so is $K^\Uparrow$.

2) \ If $K_1\subseteq K_2$, then $K_1^\Uparrow\subseteq
K_2^\Uparrow$. In other words, the operation $\Uparrow$ is
monotone.

3) \ $(K_1\cap K_2)^\Uparrow\subseteq K_1^\Uparrow\cap
K_2^\Uparrow$; this inclusion can be strict.

4) \ $K_1^\Uparrow\cup K_2^\Uparrow\subseteq(K_1\cup
K_2)^\Uparrow$; this inclusion can be strict.
\end{Prop}
\begin{proof}\

1) Since $K^\Uparrow$ is $\ld$-controllable,
$\overline{K^\Uparrow}$ is also $\ld$-controllable by 3) of Remark
\ref{Pcol}. Therefore, we get that $\overline{K^\Uparrow}\subseteq
K^\Uparrow$. The converse inclusion is always true, so
$K^\Uparrow$ is prefix closed.

2) It follows immediately from the definition of the operation
$\Uparrow$.

3) The first part follows directly from 2). The example presented
in the proof of 2) of Proposition \ref{PContr} shows us that this
inclusion can be strict.

4) It follows from 2) that the first part holds. For the second
part, let us see the following example: Keep $L(G),E_{uc}$, and
$d$ in Example \ref{Eclo}. Let $K_1=\{\epsilon,a,af\}$,
$K_2=\{\epsilon,a,ag\}$, and $\ld=0.0025$. Then by definition we
have that $K_1^\Uparrow\cup K_2^\Uparrow=\{\epsilon\}$. However,
$(K_1\cup K_2)^\Uparrow=\{\epsilon,a,ag,af\}$, so the inclusion
may be strict.
\end{proof}

\vspace{0.3cm}Recall that Proposition \ref{PRea} tells us that
$B(K,\ld)^\uparrow$ is the supremal realization of
$\ld$-controllable language $K$. In fact, this result can be
generalized to the case that $K$ is not necessarily
$\ld$-controllable.
\begin{Thm}Let $K$ be a language. Then $B(K,\ld)^\uparrow$ is the supremal realization of $K^\Uparrow$.
\end{Thm}
\begin{proof}By Proposition \ref{PRea}, we know that $B(K^\Uparrow,\ld)^\uparrow$ is the supremal realization
of $K^\Uparrow$, so it is sufficient to show that
$B(K,\ld)^\uparrow=B(K^\Uparrow,\ld)^\uparrow$. Note that
$K^\Uparrow\subseteq K$ and the operation $\uparrow$ is monotone,
therefore we have that $B(K^\Uparrow,\ld)\subseteq B(K,\ld)$, and
furthermore, $B(K^\Uparrow,\ld)^\uparrow\subseteq
B(K,\ld)^\uparrow$. For the converse inclusion, set $K'=\{t\in
K:\exists s\in B(K,\ld)^\uparrow\mbox{ such that
}d(s,t)\leq\ld\}$, and then observe that
$d(B(K,\ld)^\uparrow,K')\leq\ld$. Therefore, $K'$ is
$\ld$-controllable, and $B(K,\ld)^\uparrow$ is a realization of
$K'$. Since $B(K',\ld)^\uparrow$ is the supremal realization of
$K'$ by Proposition \ref{PRea}, we have that
$B(K,\ld)^\uparrow\subseteq B(K',\ld)^\uparrow$. By the previous
arguments, we know that $K'\in\mathscr{C}_\ld(K)$. This means that
$K'\subseteq K^\Uparrow$, and moreover,
$B(K',\ld)^\uparrow\subseteq B(K^\Uparrow,\ld)^\uparrow$.
Consequently, $B(K,\ld)^\uparrow\subseteq
B(K^\Uparrow,\ld)^\uparrow$, as desired.
\end{proof}

\vspace{0.3cm}From the proof of the above theorem, we can easily
get the following:
\begin{Coro} $B(K,\ld)^\uparrow=\nobreak B(K^\Uparrow,\ld)^\uparrow=\nobreak\cup_{M\in\mathscr{C}_\ld(K)}B(M,\ld)^\uparrow$.
\end{Coro}

\section{Optimality of Realizations}
By introducing metric to the set of languages, we have defined the
realization of a language $K$ as the controllable language that is
similar to $K$. Though there is an index $\ld$ reflecting the
similarity, the elements of two similar languages may be quite
different from each other. It is comprehensible since the
similarity characterized by a metric yields that two elements are
not identical unless the distance between them is $0$.

In view of supervisory control, we are interested in finding a
realization of $K$ that has common elements with $K$ as many as
possible on the one hand and has different elements with $K$ as
few as possible on the other hand.  To this end, let us consider
the following problem.

{\it Optimal Control Problem (OCP):} Given $\ld\in[0,1]$ and a
nonempty language $K\subseteq L(G)$, find a supervisor $S$ such
that:
\begin{enumerate}
    \item $d(L(S/G),K)\leq\ld$.
    \item $L(S/G)$ is {\it Pareto optimal} with respect to the following two sets which serve as
    measure of performance:
\begin{itemize}
        \item The {\it common element measure} of $S$, $CEM(S)$, defined as
        $CEM(S)=L(S/G)\cap K.$
        \item The {\it different element measure} of $S$, $DEM(S)$, defined as
        $DEM(S)=L(S/G)\backslash K.$
    \end{itemize}
\end{enumerate}
{\it Pareto optimality} means that any possible improvement of
$CEM(S)$ by enlarging this set is necessarily accompanied by an
enlargement of $DEM(S)$. Similarly, any possible improvement of
$DEM(S)$ by reducing this set is necessarily accompanied by a
reduction of $CEM(S)$.

For simplicity, we suppose that the language $K$ is prefix closed
in this section. Let us first describe two extreme solutions to
OCP.

\begin{Thm}\

1) \ OCP has a solution satisfying $CEM(S)=K$ if and only if
$K^\downarrow\subseteq B(K,\ld)$.

2) \ OCP has a solution satisfying $DEM(S)=\emptyset$ if and only
if $d(K^\uparrow,K)\leq\ld$.
\end{Thm}
\begin{proof}\

1) Suppose that OCP has a solution satisfying $CEM(S)=K$. Then by
the definition of $CEM(S)$ we see that $K\subseteq L(S/G)$, which
means that $K^\downarrow\subseteq L(S/G)$. Note that $L(S/G)$ is a
realization of $K$, so $L(S/G)\subseteq B(K,\ld)^\uparrow\subseteq
B(K,\ld)$ by Proposition \ref{PRea}. Hence, $K^\downarrow\subseteq
B(K,\ld)$.

Conversely, suppose that $K^\downarrow\subseteq B(K,\ld)$. Then
there exists a supervisor $S_0$ such that
$L(S_0/G)=K^\downarrow\subseteq B(K,\ld)$ since $K^\downarrow$ is
controllable. Therefore, $d(L(S_0/G),K)\leq\ld$, and moreover,
$CEM(S_0)$\\$=L(S_0/G)\cap K=K$ and $DEM(S_0)=L(S_0/G)\backslash
K=K^\downarrow\backslash K.$ Clearly, $L(S_0/G)$ is Pareto
optimal.

2) The proof is similar to that of 1). Suppose that OCP has a
solution satisfying $DEM(S)=\emptyset$. Then by the definition of
$DEM(S)$ we have that $L(S/G)\subseteq K$. We thus get that
$L(S/G)\subseteq K^\uparrow$. Since $d(L(S/G),K)\leq\ld$, it is
obvious that $d(K^\uparrow,K)\leq\ld$.

Conversely, suppose that $d(K^\uparrow,K)\leq\ld$. Since
$K^\uparrow$ is controllable, there is a supervisor $S'$ such that
$L(S'/G)=K^\uparrow\subseteq K$. Thus we obtain that
$d(L(S'/G),K)\leq\ld$, $CEM(S')=L(S'/G)\cap K=K^\uparrow$, and
$DEM(S')=L(S'/G)\backslash K=\emptyset.$ Clearly, $L(S'/G)$ is
necessarily Pareto optimal.
\end{proof}

\vspace{0.3cm}The next theorem shows us that OCP has a solution
whenever $K$ is $\ld$-controllable. It implies that we can obtain
a Pareto optimal realization from any realization (in particular,
the supremal realization). The resultant realization will
significantly improve the original one.

\begin{Thm}OCP has a solution if and only if $K$ is $\ld$-controllable.\label{Top}
\end{Thm}

The necessity of the above theorem is obvious. For the proof of
the sufficiency, we need several lemmas. In fact, the process of
proving the sufficiency is just the process of finding a Pareto
optimal realization from a given realization.

Suppose that $\widetilde{K}$ is a realization of $K$, where
$\widetilde{K}$ is prefix closed, but not necessarily Pareto
optimal. We take two steps to find a Pareto optimal realization
from $\widetilde{K}$: 1). Find a realization $\widetilde{K}_s$ by
improving $\widetilde{K}$ such that $\widetilde{K}_s\cap
K\supseteq\widetilde{K}\cap K$ and $\widetilde{K}_s\cap K$ is as
large as possible, which helps us find more common elements; 2).
Find a realization $N_m$ by improving $\widetilde{K}_s$ such that
$\widetilde{K}_s\cap K\subseteq N_m\subseteq\widetilde{K}_s$ and
$N_m$ is as small as possible, which helps us reduce the different
elements.

For the Step 1), define $\mathscr{M}=\{M:M\subseteq
K\backslash\widetilde{K}\mbox{ and }M^\downarrow\subseteq
K\cup\widetilde{K}\}.$ Observe that $\emptyset\in\mathscr{M}$, so
the class is not empty. Moreover, $\mathscr{M}$ is closed under
arbitrary unions, hence it contains a unique supremal element,
denoted $M_s$, with respect to set inclusion. Clearly,
$M_s=\cup_{M\in\mathscr{M}}M$. The following lemma which is
analogous to Lemma 5.1 in \cite{CL91} provides some
characterizations of $M_s$.

\begin{Lem}\

1) \ $M_s=M_s^\downarrow\cap(K\backslash\widetilde{K})
=(K\cup\widetilde{K})^\uparrow\cap(K\backslash\widetilde{K})$.

2) \ $\widetilde{K}\cup M_s^\downarrow=\widetilde{K}\cup
M_s$.\label{Lcom}
\end{Lem}
\begin{proof}\

1) We first prove the first equality. Obviously, $M_s\subseteq
M_s^\downarrow\cap(K\backslash\widetilde{K})$. Conversely, write
$N$ for $M_s^\downarrow\cap(K\backslash\widetilde{K})$. Then
$N\subseteq M_s^\downarrow$ and $N\subseteq
K\backslash\widetilde{K}$. The former means that
$N^\downarrow\subseteq M_s^\downarrow\subseteq
K\cup\widetilde{K}$. Consequently, $N\in\mathscr{M}$, and thus
$N\subseteq M_s$. So
$M_s=M_s^\downarrow\cap(K\backslash\widetilde{K})$. The second
equality can be verified in a similar may, so we omit the proof.

2) Using 1), we get that
\begin{eqnarray*} \widetilde{K}\cup M_s
 &=&\widetilde{K}\cup(M_s^\downarrow\cap(K\backslash\widetilde{K}))\\
 &=&(\widetilde{K}\cup M_s^\downarrow)\cap(\widetilde{K}\cup(K\backslash\widetilde{K}))\\
 &=&(\widetilde{K}\cup M_s^\downarrow)\cap(K\cup\widetilde{K})\\
 &=&\widetilde{K}\cup M_s^\downarrow,
 \end{eqnarray*}that is, $\widetilde{K}\cup M_s^\downarrow=\widetilde{K}\cup M_s$.
\end{proof}

\vspace{0.3cm}
The next proposition shows that by adding $M_s$ to
$\widetilde{K}$, we can get a better realization in the sense that
the common elements may be improved without worsening the
different elements.
\begin{Prop}Let $\widetilde{K}_s=\widetilde{K}\cup M_s$. Then $\widetilde{K}_s$ is a realization of
$K$; moreover, $\widetilde{K}_s=\overline{\widetilde{K}_s}$,
$\widetilde{K}_s\cap K\supseteq\widetilde{K}\cap K$ and
$\widetilde{K}_s\backslash K=\widetilde{K}\backslash
K$.\label{Pcom}
\end{Prop}
\begin{proof}From 2) of Lemma \ref{Lcom}, we see
that $\widetilde{K}_s$ is controllable. It is clear that
$d(\widetilde{K}_s,K)\leq\ld$ since $d(\widetilde{K},K)\leq\ld$
and $M_s\subseteq K$. Therefore, $\widetilde{K}_s$ is a
realization of $K$. The remainder of this proposition follows
readily from Lemma \ref{Lcom}.
\end{proof}

\vspace{0.3cm} For Step 2), we require the following fact.
\begin{Lem}Let $s\in E^*$, and suppose that there is a chain of prefix closed languages over $E$: $$X_1\supseteq
X_2\supseteq\cdots\supseteq X_i\supseteq\cdots$$satisfying that
$d(s,X_i)\leq\ld$ for all $i$.
 Then $d(s,\cap_i X_i)\leq\ld$.\label{Lbound}
\end{Lem}
\begin{proof}If the length of the chain is finite, then the lemma holds evidently. We now prove the case
that the length of the chain is infinite. Set $B_i=X_i\cap
B(s,\ld)$ for all $i$. Since $d(s,X_i)\leq\ld$, we know  by Lemma
\ref{Lmin} that $\min\limits_{x_i\in X_i}d(s,x_i)\leq\ld$. So
there is $x_i\in X_i$ with $d(s,x_i)\leq\ld$. Thus $B_i$ is not
empty and there is a chain: $$B_1\supseteq
B_2\supseteq\cdots\supseteq B_i\supseteq\cdots.$$ By
contradiction, assume that $d(s,\cap_iX_i)>\ld$. Then again by
Lemma \ref{Lmin} there is no $x$ in $\cap_iX_i$ such that
$d(s,x)\leq\ld$. This means that for any $x_i\in B_i$, there
exists $j>i$ such that $x_i\not\in X_j$, that is, $x_i\not\in
B_j$. In particular, we now take $b_1\in B_1$ with the minimal
length (i.e., $l(b'_1)\geq l(b_1)$ for any $b'_1\in B_1$). Then
there exists $j_1>1$ such that $b_1\not\in B_{j_1}$. Next, note
that $B_{j_1}\neq\emptyset$, and take $b_2\in B_{j_1}$ with the
minimal length. Clearly, $l(b_2)\geq l(b_1)$ since
$B_{j_1}\subseteq B_1$. By the same token, we have $b_{r+1}\in
B_{j_r}$, $r=1,2,\cdots$, such that $l(b_{r+1})\geq l(b_r)$ and
$l(b_{r+1})$ is the minimal length of strings in $B_{j_r}$.
Because the set $\{b\in E^*:l(b)\leq l(s)\}$ is finite, there is
$r_0$ such that $l(b_{r_0+1})>l(s)$. Let $b'$ be the prefix of
$b_{r_0+1}$ with length $l(s)$. Then we see by the definition of
Baire metric that $d(s,b')<d(s,b_{r_0+1})\leq\ld$. Since
$X_{j_{r_0}}$ is prefix closed, we get that $b'\in X_{j_{r_0}}$,
and thus $b'\in B_{j_{r_0}}$. Because $l(b_{r_0+1})$ is the
minimal length of strings in $B_{j_{r_0}}$, we have that
$l(b_{r_0+1})\leq l(b')$, which contradicts the previous argument
that $l(b_{r_0+1})>l(s)=l(b')$. This completes the proof of the
lemma.
\end{proof}

\vspace{0.3cm} Let us now define
$$\mathscr{N}=\{N:\widetilde{K}_s\cap K\subseteq
N\subseteq\widetilde{K}_s,N=\overline{N}, \mbox{ and }N\mbox{ is a
realization of } K\}.$$ This class is not empty since
$\widetilde{K}_s\in\mathscr{N}$ by Proposition \ref{Pcom}. Recall
that the intersection of two realizations of a $\ld$-controllable
language does not necessarily give a realization, so the class
$\mathscr{N}$ has no infimal element in general. Nevertheless, we
have the following result.

\begin{Lem}
The class $\mathscr{N}$ has a minimal element with respect to set
inclusion.\label{Lm}
\end{Lem}
\begin{proof}Clearly, $(\mathscr{N},\supseteq)$ is a partially ordered set. If each chain in
$(\mathscr{N},\supseteq)$ has a lower bound, then by Zorn's Lemma
there is a minimal element of $\mathscr{N}$. So it suffices to
show that any chain in $(\mathscr{N},\supseteq)$ has a lower
bound. Let $$N_1\supseteq N_2\supseteq\cdots\supseteq
N_i\supseteq\cdots$$ be a chain in $(\mathscr{N},\supseteq)$. Then
we have that $\widetilde{K}_s\cap K\subseteq\cap_i
N_i\subseteq\widetilde{K}_s$ and $\cap_i N_i=\overline{\cap_i
N_i}$. By Lemma \ref{Lbound}, we see that $d(s,\cap_i N_i)\leq\ld$
for any $s\in K$. On the other hand, $d(t,K)\leq\ld$ for any
$t\in\cap_i N_i$ since $\cap_i N_i\subseteq\widetilde{K}_s$ and
$d(\widetilde{K}_s,K)\leq\ld$. As a result, $d(\cap_i
N_i,K)\leq\ld$. Note that $\cap_i N_i$ is controllable, so $\cap_i
N_i$ is a realization of $K$. Thus $\cap_i N_i\in\mathscr{N}$ and
$\cap_i N_i$ is a lower bound of the chain. This finishes the
proof.
\end{proof}

\vspace{0.3cm}
Based on the previous lemmas, we can now prove the main result of
this section.

{\it Proof of Theorem \ref{Top}:} It remains to prove the
sufficiency. Suppose that $K$ is $\ld$-controllable and
$\widetilde{K}$ is a realization of $K$. Since $K$ has been
assumed to be prefix closed, we know that
$\overline{\widetilde{K}}$ is a realization of $K$ by the proof
for 3) of Remark \ref{Pcol}. For convenience of notation, we write
$\widetilde{K}$ for $\overline{\widetilde{K}}$. It follows from
Lemma \ref{Lm} that $\mathscr{N}$ has a minimal element, say,
$N_m$. We claim that $N_m$ is a solution to OCP.

By the definition of $\mathscr{N}$, we know that $N_m$ is a
realization of $K$. That is, there exists a supervisor $S_m$ such
that $L(S_m/G)=N_m$ and $d(L(S_m/G),K)\leq\ld$. It remains to
verify that $N_m$ is Pareto optimal. Seeking a contradiction,
suppose that there is another realization $N'$, which is prefix
closed, of $K$ satisfying the following:
\begin{description}
    \item[(1)] $N'\cap K \supseteq N_m\cap K$;
    \item[(2)] $N'\backslash K\subseteq N_m\backslash K$; and
    \item[(3)] at least one of the two inclusions above is strict.
\end{description}

Observe first that $N'\subseteq K\cup\widetilde{K}$. Otherwise,
there exists $s\in N'$, but $s\not\in K\cup\widetilde{K}$. We thus
see that $s\not\in K$, which means that $s\in N_m$ by (2). But
$N_m\subseteq\widetilde{K}_s=\widetilde{K}\cup M_s\subseteq
K\cup\widetilde{K}$ by Lemma \ref{Lcom}. This contradicts with
$s\not\in K\cup\widetilde{K}$. Since $\widetilde{K}_s\cap
K\subseteq N_m$, we get by (1) that $\widetilde{K}_s\cap
K\subseteq N_m\cap K\subseteq N'\cap K\subseteq N'$, namely,
$\widetilde{K}_s\cap K\subseteq N'$.

Set $M'=N'\backslash\widetilde{K}$. From $N'\subseteq
K\cup\widetilde{K}$, we find that $M'\subseteq
K\backslash\widetilde{K}$ and $M'^\downarrow\subseteq
K\cup\widetilde{K}$. Therefore $M'\in\mathscr{M}$, and thus
$M'\subseteq M_s$. Notice that $M_s\subseteq(\widetilde{K}\cup
M_s)\cap K=\widetilde{K}_s\cap K\subseteq N_m$. Hence $M'\subseteq
N_m$ and $K\cap\widetilde{K}\subseteq N_m$. For any $s\in
N'\subseteq K\cup\widetilde{K}$, if $s\not\in K$, then by (2) we
have that $s\in N_m$; if $s\not\in\widetilde{K}$, then $s\in M'$
and by the previous argument $M'\subseteq N_m$ we also have that
$s\in N_m$; if $s\in K\cap\widetilde{K}$, we still have that $s\in
N_m$ since $K\cap\widetilde{K}\subseteq N_m$. Consequently,
$N'\subseteq N_m\subseteq\widetilde{K}_s$. This, together with the
proven fact $\widetilde{K}_s\cap K\subseteq N'$, yields that
$N'\in\mathscr{N}$. Then we get $N'=N_m$ since $N_m$ is a minimal
element. It forces that neither of the inclusions in (1) and (2)
is strict, which contradicts with (3). Therefore, $N_m$ is Pareto
optimal, finishing the proof of the theorem. \hfill$\blacksquare$

\vspace{0.3cm}We give a simple example to illustrate the process
of finding a  Pareto optimal realization from any given
realization.

\begin{Eg}Let $E=\{a,b,c,e,f,u\}$ and $E_{uc}=\{u\}$. The automaton $G$ that generates $L(G)$ is
depicted in Figure 1. Let $K=\{\epsilon,a,ab,ac\}$ and
$\ld=\frac{1}{16}$. The metric $d$ on $E$ is defined as follows:
\begin{displaymath}
d(x,y)=\left\{ \begin{array}{ll} 0, &\textrm{if $x=y$}\\
\frac{1}{4}, & \textrm{if $(x,y)\in\{(b,e),(e,b),(c,f),(f,c),(b,f),(f,b)\}$}\\
\frac{1}{2}, & \textrm{otherwise.}
\end{array} \right.
\end{displaymath}
\begin{figure}\caption{Automaton of Example 2.}
\includegraphics{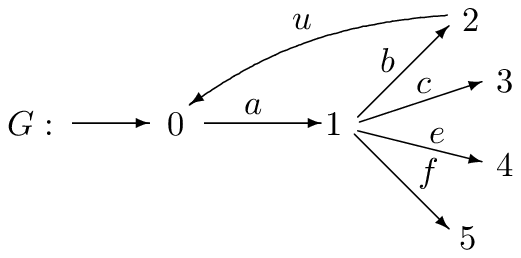}
\end{figure}
\end{Eg}Observe that $K$ is not controllable with respect to $L(G)$ and $E_{uc}$, but it is
$\frac{1}{16}$-controllable with respect to $L(G)$ and $E_{uc}$.
It is easily verified that $\widetilde{K}=\{\epsilon,a,ae,af\}$
can serve as a realization. Such a realization is not, however,
Pareto optimal since we can enlarge $CEM(S)$ (without enlargement
of $DEM(S)$) by adding $ac$ to $\widetilde{K}$, or reduce $DEM(S)$
(without reduction of $CEM(S)$) by removing $ae$ from
$\widetilde{K}$.

We are now ready to use the procedure in the proof of Theorem
\ref{Top}  to obtain a Pareto optimal realization from
$\widetilde{K}$.

Keep the previous notation of this section. By an easy
calculation, we get that the supremal element of
$$\mathscr{M}=\{M:M\subseteq K\backslash\widetilde{K}\mbox{ and }M^\downarrow\subseteq
K\cup\widetilde{K}\}$$ is $\{ac\}$, and thus
$\widetilde{K}_s=\widetilde{K}\cup M_s=\{\epsilon,a,ac,ae,af\}$.
Further, we have that
\begin{eqnarray*}
\mathscr{N}
 &=&\{N:\widetilde{K}_s\cap K\subseteq N\subseteq\widetilde{K}_s,N=\overline{N}, \mbox{ and }N\mbox{
is a realization of } K\}\\
 &=&\{\{\epsilon,a,ac,ae\},\{\epsilon,a,ac,af\},\{\epsilon,a,ac,ae,af\}\}.
\end{eqnarray*} Observe that
$\mathscr{N}$ has two minimal elements: $\{\epsilon,a,ac,ae\}$ and
$\{\epsilon,a,ac,af\}$. They give rise to two Pareto optimal
realizations of $K$.

\section{Illustrative Example}
In this section, we apply the notion of $\ld$-controllability to a
machine which is a modified version of that studied in
\cite{KGM93}. Then we further explain the necessity of optimal
control and illustrate the process of finding a Pareto optimal
realization from an arbitrary realization.
\begin{figure}
\caption{Automaton $G$ of Section V.}
\includegraphics{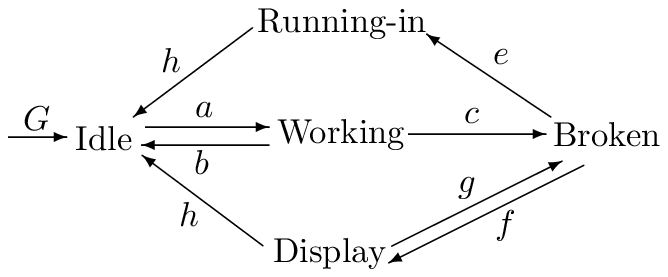}
\end{figure}

The plant $G$, shown in Fig. 1, is a machine consisting of five
states: {\it Idle, Working, Broken, Display,} and
\textit{Running-in}. The events of the plant model are listed in
Table 1.

\begin{tabular}{c l}
  \multicolumn{2}{l}{Table 1. Meaning of events.}\\
  \hline
  Event & \ \ Event description \\
  \hline
  a & \ \ start (Controllable) \\
  b & \ \ stop (Controllable) \\
  c & \ \ fail (Uncontrollable) \\
  e & \ \ replace (Controllable) \\
  f & \ \ repair (Controllable) \\
  g & \ \ reject (Uncontrollable) \\
  h & \ \ approve (Uncontrollable) \\
  \hline
  \end{tabular}
\vspace{0.4cm}

We suppose that the machine needs a thorough inspection after a
period of run, say three ``start" events for simplicity. Thus the
specification $K$ is only concerned with strings that contain at
most three $a$'s. More explicitly, $K$ is generated by the
automaton $H$ depicted in Fig. 2. Clearly, $K$ is not
controllable. Nevertheless, we can image that certain events such
as ``repair" and ``replace" are similar, especially after some
occurrences of ``fail" and ``reject", since one would like to
replace a component in some circumstances rather than repair it
again and again. Formally, we define a metric $d$ on $E$ as
follows:
\begin{displaymath}
d(x,y)=\left\{ \begin{array}{ll} 0, &\textrm{if $x=y$}\\
2^{-7}, & \textrm{if $(x,y)\in\{(b,c),(c,b)\}$}\\
2^{-10}, & \textrm{if $(x,y)\in\{(e,f),(f,e)\}$}\\
1, & \textrm{otherwise.}
\end{array} \right.
\end{displaymath}

Set $\ld=2^{-14}$, and we then find that $K$ is
$2^{-14}$-controllable. In fact, it is not difficult to verify
that the language $\widetilde{K}$ generated by the automaton $H'$
depicted in Fig. 3 can serve as a realization. (Of course, there
exist other realizations, for example, $B(K,\ld)^\uparrow$.) On
the other hand, we may still add some strings that belong to $K$
and do not destroy the controllability of $\widetilde{K}$ to
$\widetilde{K}$, and may also remove some strings in
$\widetilde{K}$ but not in $K$ and keep $\widetilde{K}$
controllable. Clearly, such a process of keeping $K$ most possibly
invariable is significative and necessary, and it can be
accomplished by seeking a Pareto optimal realization as follows.
\begin{figure}
\caption{Automaton $H$ of Section V.}
\includegraphics{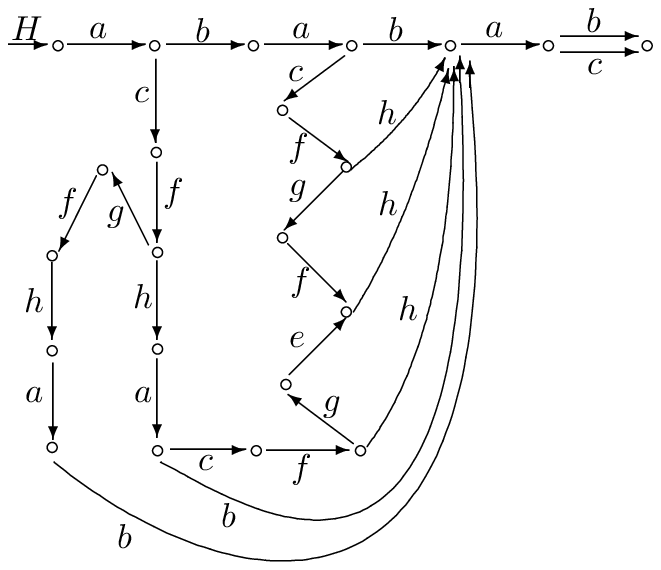}
\end{figure}
\begin{figure}
\caption{Automaton $H'$ of Section V.}
\includegraphics{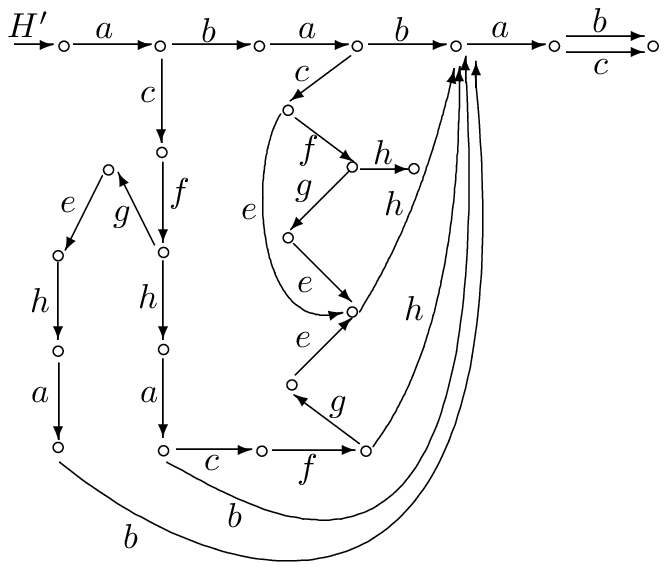}
\end{figure}

Keep the previous notation of the last section. By an easy
calculation, we get that the supremal element $M_s$ of
$\mathscr{M}=\{M:M\subseteq K\backslash\widetilde{K}\mbox{ and
}M^\downarrow\subseteq K\cup\widetilde{K}\}$ is
$\{abacfha,abacfhab,abacfhac\}$, and thus
$\widetilde{K}_s=\widetilde{K}\cup M_s$. Further, we have that
\begin{eqnarray*}
\mathscr{N}
 &=&\{N:\widetilde{K}_s\cap K\subseteq N\subseteq\widetilde{K}_s,N=\overline{N}, \mbox{ and }N\mbox{
is a realization of } K\}\\
 &=&\{\widetilde{K}_s\backslash\{abace,abaceh,abaceha,abacehab,abacehac\},\\& &
 \widetilde{K}_s\backslash\{abaceha,abacehab,abacehac\},
 \widetilde{K}_s\backslash\{abacehab\},
 \widetilde{K}_s\}.
\end{eqnarray*} Observe that
$\mathscr{N}$ has one minimal element:
$\widetilde{K}_s\backslash\{abace,abaceh,abaceha,abacehab,abacehac\}$
, which gives rise to a Pareto optimal realization of $K$. It is
worth noting that Baire metric plays a role here: although $f$ is
similar to $e$, $f$ is not allowed to be replaced by $e$ if the
machine first breaks; in other words, any realization of $K$
cannot contain the string $ace$.

\section{Conclusion and Discussion}
In this paper, we have introduced a similarity-based supervisory
control methodology for DES. By tolerating some similar behavior,
we can realize some desired behavior which is uncontrollable in
traditional SCT. A generalized notion of controllability, called
$\ld$-controllability, has been proposed. We have elaborated on
some properties of $\ld$-controllable languages and their
realizations.

There are some limits and directions in which the present work can
be extended. Note that the algorithm for $B(K,\ld)$ developed in
Section III only works for finite languages although all the
remainder results have been established for any languages. It is
desirable to find a more general algorithm. Metrics chosen here
including Baire metric and Hausdorff metric pay more attention to
the events occurring antecedently. The distance between strings or
languages that is given by such metrics may not be meaningful in
certain practical systems, and the selection of these metrics is
generally dependent on the particular problem considered. This
means that perhaps supervisory control problems based on some
other metrics or similarity measure (for example, Hamming distance
in Information Theory) need to be considered. In addition, some
other issues in standard SCT such as observability \cite{LW88} and
nonblocking \cite{RW87} remain yet to be addressed in our
framework.

\section*{Acknowledgment}
The authors would like to thank Yuan Feng for some helpful
discussions. Furthermore, the authors are very grateful to
Associate Editor Alessandro Giua and the referees for their
invaluable comments.

\end{document}